\documentclass[twocolumn,aps,epsfig,epsf]{revtex4}
\usepackage{graphicx}

\newcommand{\PD}{\partial}
\newcommand{\be}{\begin{equation}}
\newcommand{\ee}{\end{equation}}
\newcommand{\bea}{\begin{eqnarray}}
\newcommand{\eea}{\end{eqnarray}}

\begin{document}

\title{Density inhomogeneities
 in heavy ion collisions around the critical point}
\author{Kerstin Paech and Adrian Dumitru}
\affiliation{Institut f\"ur Theoretische Physik, J.W.~Goethe Universit\"at,\\
Max-von-Laue Str.\ 1, D-60438 Frankfurt am Main, Germany}

\begin{abstract}
We study the hydrodynamical expansion of a hot and baryon-dense quark
fluid coupled to classical real-time evolution of the long wavelength
modes of the chiral field. Significant density inhomogeneities develop
dynamically when the transition to the symmetry-broken state
occurs. We find that the amplitude of the density inhomogeneities is
larger for expansion trajectories crossing the line of first-order
transitions than for crossovers, which could provide some information
on the location of a critical point. A few possible
experimental signatures for inhomogeneous decoupling surfaces are
mentioned briefly.
\end{abstract}

\maketitle

\section{Introduction} 

Heavy ion collisions at high energies produce hot and baryon-dense
strongly interacting matter and so
provide the opportunity to explore the phase diagram of 
QCD~\cite{HIPhaseTrans}. 
Recent lattice QCD calculations at finite
baryon-chemical potential~\cite{Fodor:2004nz} indicate that at sufficiently
large baryon density a line of first
order transitions exists in the plane of temperature $T$ versus
baryon-chemical potential $\mu_B$. This line separates the region
where chiral symmetry is broken (as in vacuum) from that where it is
approximately restored. Moving counter-clockwise along this phase
boundary, i.e.\ towards higher $T$ and lower $\mu_B$, results in
weaker first-order transitions and finally the line of first order
transition ends at a second-order critical point~\cite{SRS}.
Simulations with semi-realistic quark masses locate the endpoint at
$T_E \simeq 160$~MeV, $\mu_{B,E}\simeq 360$~MeV. For $\mu_B<\mu_{B,E}$ no
phase transition in the strict sense occurs. Rather, the low- and
high-temperature phases are continuously connected by a rapid crossover.

Our goal here is to analyze the homogeneity of the ``fluid'' of QCD
matter as it expands and cools. In particular, we shall study
expansion trajectories passing on either side of the critical point
(i.e.\ either crossover or first order phase transition). As we will
show, in the vicinity of the critical point the expanding fluid develops
significant inhomogeneities. Such density perturbations should also be present
on the decoupling surface of hadrons. This is normally neglected
in hydrodynamical simulations of heavy-ion collisions, which
commonly assume that the hadrons freeze out at a fixed temperature or density.

The fact that decoupling surfaces are typically {\em not} homogeneous
is well known. Take, for example, the WMAP data on the temperature
fluctuations of the cosmic microwave background (CMB)~\cite{wmap}. The
background photons exhibit temperature fluctuations on the order of
$\Delta T/T\simeq 10^{-5}$. From the CMB multipoles one hopes to gather
information on their primordial origin.
In heavy-ion collisions, on the other hand, we might get hints about the
QCD phase transition, if it occurs shortly before decoupling.
The basic idea is that if a phase transition occurs then it might
leave imprints on the (energy-) density distribution on the freeze-out
hypersurface. In particular, we expect the inhomogeneities
to be smaller for crossovers and stronger for first order
transitions.

The first source of spatial density inhomogeneities in
heavy-ion collisions is due to
fluctuations in the number of participants and the number of
collisions among the beam
nucleons. Those fluctuations lead to an inhomogeneous deposition of
energy and of baryon number at central
rapidity. From the event generator UrQMD, energy
density inhomogeneities on the order of $\Delta e\sim 1$~GeV/fm$^3$ 
have been predicted for central Pb+Pb 
collisions at top SPS energy ($\surd s\simeq17 A$~GeV) which originate
from fluctuations in ``soft'' (low momentum transfer)
interactions; see fig.~4 in~\cite{Bleicher:wd}.
At RHIC energies ($\surd s\simeq100-200 A$~GeV) $\Delta e$ could increase by up
to an order of magnitude due to the additional semi-hard (``minijet'')
component, as discussed in ref.~\cite{GyRiZ}.
These authors also discussed the evolution of such initial-state
inhomogeneities using equilibrium hydrodynamics in the ideal fluid
approximation. They find that even the huge initial perturbations
predicted by the minijet model are strongly washed out until freeze
out by the hydrodynamic expansion of the hot matter. Qualitatively,
this can be understood from the following simple argument. The radius
$L\simeq1$~fm of a hotspot grows linearly in time while its density
drops inversely proportional to its volume so that $\Delta\rho\sim
1/t^3$ for expansion in three dimensions. The overall duration of the
hydrodynamic expansion is expected to be (at least) on the order of
the radius of the colliding nuclei~\cite{HIPhaseTrans,HK}, roughly $\sim 5L$.
Therefore, any
initial density concentration should be diluted by about a
factor of $100$. Hence, initial perturbations of order one would
leave traces on the percent level only, at the time of decoupling. 
For other studies of early-stage density inhomogeneities and their
hydrodynamic evolution
see refs.~\cite{Drescher:2000ec} and~\cite{Socolowski:2004hw},
respectively~\cite{comment}.

Here, we discuss a different source of density inhomogeneities, namely
those possibly generated in the course of a non-equilibrium transition
from a (nearly) chirally symmetric state at high temperature and
density to the state of broken symmetry at decoupling. The rather
rapid transition expected to occur in high-energy heavy-ion collisions
very likely forces the long wavelength modes of the
chiral condensate out of equilibrium. This should then reflect in a
rather non-uniform distribution of energy and baryon density in space
(or, more precisely, on the decoupling hypersurface).
Most notably, since ``freeze out'' (decoupling of all particles) in
heavy-ion collisions occurs shortly after the transition to the
broken phase, such perturbations generated in the late stages of the
evolution could largely survive and leave detectable traces in the
final state. In this regard, we also recall the results of
ref.~\cite{Bot Aichelin} who employed the collisionless Vlasov
equation to study the real-time evolution of small initial density
fluctuations within the NJL model. They observed an increase of
the fluctuations already for the case where the expectation value of
the chiral condensate was fixed to its equilibrium value, cf.\ their
eq.~(2). Here, we also treat the non-equilibrium dynamics of the
chiral condensate, cf.\ our
eq.~(\ref{EulerLagrange}) below, which probes the structure of the
effective potential in the vicinity of the critical point.

\section{The Model} 
For the current
studies we extend the model from ref.~\cite{Paech:2003fe} to allow for
nonvanishing baryon density $\rho$.  The Gell-Mann-Levy Lagrangian~\cite{GML}
\bea {\cal L} &=&
 \overline{q}\,\left[i\gamma ^{\mu}\partial _{\mu}-g(\sigma +\gamma _{5}
 \vec{\tau} \cdot \vec{\pi} )\right]\, q\nonumber\\
&+& \frac{1}{2}\left (\partial _{\mu}\sigma\right)^2 + \frac{1}{2}\left(
\partial _{\mu} \vec{\pi} \right)^2 - U(\sigma ,\vec{\pi})\quad.
\label{sigma}
\eea 
provides an effective theory for chiral symmetry breaking in QCD.
It describes the
interaction of two flavors of constituent quarks $q=(u,d)$ with the
chiral field $\phi_a = (\sigma,\vec{\pi})$. The potential, which exhibits
both spontaneously and explicitly broken chiral symmetry, is
\begin{equation} \label{T=0_potential} 
U(\sigma ,\vec{\pi} )=\frac{\lambda
^{2}}{4}(\sigma ^{2}+\vec{\pi} ^{2} - {\it v}^{2})^{2}-h_q\sigma
-U_0\quad. 
\end{equation} 
The vacuum expectation values of the condensates
are $\langle\sigma\rangle ={\it f}_{\pi}$ and $\langle\vec{\pi}\rangle
=0$, where ${\it f}_{\pi}=93$~MeV is the pion decay constant.
The explicit symmetry breaking term is due to the 
non-zero pion mass,
$h_q=f_{\pi}m_{\pi}^{2}$, where $m_{\pi}=138$~MeV.  
This leads to $v^{2}=f^{2}_{\pi}-{m^{2}_{\pi}}/{\lambda ^{2}}$.  The value of 
$\lambda^2 = 20$ leads to a $\sigma$-mass, 
$m^2_\sigma=2 \lambda^{2}f^{2}_{\pi}+m^{2}_{\pi}$, approximately
equal to 600~MeV.

We assume that the quarks constitute a thermalized fluid, which
provides an expanding background in which the long-wavelength modes of
the chiral condensate evolve. Integrating out the quarks generates an
effective potential for $\phi_a$; computing to one loop and for a
homogeneous background (on the scale of a ``fluid element''), this
contribution is given by
\bea 
V_{\rm eff}(\phi,T,\mu) &=&
U(\phi)  - d_{q}\; T \int \! \frac{{\rm d}^3p}{(2\pi)^3} \\ 
&& \left\{ \log
\left(1+e^{{(\mu-E)}/{T}} \right) + \left( \mu \rightarrow - \mu \right)
\right\}~.\nonumber \label{effpot} 
\eea 
Here, $d_q=12$ denotes the color-spin-isospin degeneracy of the
quarks and $\mu=\mu_B/3$ the quark-chemical potential. The two terms
inside the integral correspond to the thermal contributions of quarks and
anti-quarks, respectively; a (divergent) vacuum contribution has been
absorbed into the $T$ and $\mu$ independent potential $U$.
$V_{\rm eff}$ depends on the order
parameter field through the effective mass of the quarks, $m_q^2=g^2\phi^2$,
which enters the expression for the single-particle energy
$E=\sqrt{{\vec{p}}^{\;2}+m^2_q}$.

For sufficiently small quark-chemical potential $\mu$ one finds a
smooth transition to approximately massless quarks at high $T$.
For larger chemical potential, however, the effective potential
exhibits a first-order phase transition~\cite{SMMR}. Along
the line of first-order transitions the effective potential exhibits
two degenerate minima which are seperated by a
``nucleation barrier''. This barrier decreases with
$\mu$ and the
two minima approach each other. At $\mu_E$, finally, the barrier vanishes,
and so does the latent heat. For $g=3.3$, which leads to a constituent
quark mass in vacuum of $\approx 307$~MeV, the second-order critical
point is located at $T_E\approx100$~MeV, $\mu_E\approx200$~MeV. Increasing
the quark-field coupling $g$ moves the endpoint $E$ towards the
temperature axis~\cite{ove}  ($\mu_E$ becomes =0 at about
$g\approx3.7$~\cite{Paech:2003fe}) and to slightly higher temperature. In
what follows, we fix $g=3.3$.

The location of the endpoint does not agree quantitatively with that from
recent lattice QCD studies with realistic quark masses, which
find $T_E\approx 160$~MeV and $\mu_E\approx
120$~MeV~\cite{Fodor:2004nz}. This failure
is common to several models, c.f.\ fig.~6 in~\cite{stephanov} and
could be due to the neglect of heavier resonance states in the above
effective Lagrangian, see e.g.\ the discussion in~\cite{det}.
Also, if deconfinement and chiral symmetry restoration occur
simultaneously then the energy density contributed by Polyakov loops
in the deconfined phase should be included as well~\cite{Ploops}.
Finally, note that the model fails to describe the nuclear matter
ground state, which has non-zero pressure and is located in the phase
coexistence region (Fig.~\ref{e_rho_all} below).
Nevertheless, one might hope that,
qualitatively, the dynamics of relativistic quark fluids near the
endpoint is not affected by the deficiencies of this simple model.

The classical equations of motion for the chiral fields are
\be
\partial_{\mu}\partial^{\mu}\phi_a + \frac{\delta 
V_{\rm eff}}{\delta\phi_a}=0~.
\label{EulerLagrange}
\ee
We do not account explicitly for a damping term due to decay
processes or elastic
collisions of the particles forming the condensate~\cite{dissip}. An
ensemble average over random initial field configurations implicitly
introduces such effects at the classical level. The
(in-)accuracy of this approximation should be a matter of further
study~\cite{damping}. In~(\ref{EulerLagrange}),
 the only explicit damping of field oscillations
arises from the expansion of the fireball.

The dynamical
evolution of the thermalized degrees of freedom (fluid of quarks) is
determined by the conservation laws for energy, momentum and (net)
baryon charge:
\bea \label{contEq_Q} \PD_\mu \left(
T^{\mu\nu}_{\rm fluid} + T^{\mu\nu}_\phi \right) &=& 0\nonumber\\ 
\PD_\mu \left(\rho u^\mu\right) &=& 0~.
\eea 
Here, $u^\mu$ is the fluid four-velocity and $T^{\mu\nu}_{\rm fluid}$
its energy-momentum tensor, which we assume to be of
perfect fluid form. $T^{\mu\nu}_\phi$, in turn, is the energy-momentum
tensor of the classical fields which can be obtained from the above
Lagrangian in the standard fashion~\cite{Paech:2003fe,Mishustin:1997iz}. Note
that we do not assume that the chiral fields are equilibrated with the
heat bath of quarks. Hence, the fluid pressure depends not only on the
energy and baryon density in the local rest frame but also on the
chiral (order-parameter) field, i.e.~$p=p(e,\rho,\phi)$.  

We employ eq.~(\ref{EulerLagrange}) to also propagate initial field
fluctuations through the transition; that is,
our initial condition includes some generic ``primordial'' spectrum of
fluctuations (see below) which then evolve in the effective potential
generated by the matter fields.

\section{Results} \subsection{Initial Conditions}

We employ the following set of simple initial conditions to illustrate
qualitative effects. At $t=0$, we initialize a sphere of
hot and dense quarks with radius $R=$~5 fm and no initial
collective motion, $\vec{v}(t=0)=0$. The energy and baryon-density
distribution is taken as
\bea 
e(t=0,\vec{x}) &=& \frac{e_{\rm eq}}{ 1 +
\exp\left(\frac{r-R}{a}\right)} \nonumber\\ 
\rho(t=0,\vec{x}) &=&
\frac{\rho_{\rm eq}}{ 1 + \exp\left(\frac{r-R}{a}\right)}~, \label{DensProf}
\eea 
with a surface thickness of $a=0.3$~fm.

Within that sphere, the average 
chiral field corresponds to the minimum of $V_{\rm eff}(e_{\rm eq},
\rho_{\rm eq})$. Specifically, we choose
\bea 
\sigma(t=0,\vec{x})&=&\delta\sigma(\vec{x}) + f_\pi +\frac{
\sigma_{\rm eq} - f_\pi}{ 1 + \exp\left(\frac{r-R}{a}\right)} \nonumber\\
&=& \delta\sigma(r,\varphi,\theta) + \langle \sigma \rangle(r)
\nonumber\\
\vec{\pi}(t=0,\vec{x})&=& \delta\vec\pi~,
\eea 
with $\sigma_{\rm eq}\approx0$ the expectation
value of the $\sigma$ field corresponding to $e_{\rm eq}$ and $\rho_{\rm
eq}$. Thus, the chiral condensate nearly vanishes at the center, where the
energy density of the quarks is large, and then quickly interpolates to
$f_\pi$ where the matter density is low.
The system subsequently expands hydrodynamically on account of the
nonzero pressure. 

$\delta\sigma(\vec{x})$ represents Gaussian random fluctuations of the
fields which are distributed according to 
\be \label{FieldFlucs}
P[\delta\phi_a] \propto
\exp\left( - \delta\phi_a^2/2\left<\delta\phi_a^2\right>\right) \, . 
\ee
The results presented here were obtained with a width of
$\surd\langle\delta\sigma^2\rangle = v/3$,
$\surd\langle\delta\vec\pi^2\rangle  = 0$.
These relatively moderate amplitudes suffice to probe the structure of
the effective potential near the transition. Of course, larger
fluctuations would amplify the effects shown below.
We correlate the initial field fluctuations over approximately
$1$~fm as described in~\cite{Paech:2003fe}. Our focus is on how those
``primordial'' fluctuations evolve through the various transitions.

For definiteness, we shall consider two different sets of initial conditions: 
for set (I) we start the evolution at fixed
initial energy density $e_{\rm eq}= 2.8e_0$ but {\em vary}
the initial baryon density \mbox{$\rho_{\rm eq} = (0, 0.6, 1.6, 2.1,
  2.4, 2.8)\rho_0$}; for set
(II), on the other hand, we start at fixed
initial net baryon density $\rho_{\rm eq}=1.7\rho_0$
but {\em vary} the initial energy density
\mbox{$e_{\rm eq} = (1.4, 1.9, 2.9)e_0$}. Here, $e_0$ and $\rho_0$
denote nuclear matter ground state energy and baryon density, respectively.
For low baryon density (I) (high energy density (II)), the expansion will then
proceed through a crossover, while a baryon dense (I) (energy dilute (II)) 
droplet will decay via a first-order phase transition. (In contrast,
in~\cite{Paech:2003fe,ove} the type
and strength of the transition was controlled via the coupling
constant $g$ rather than the baryon density.)
Our goal is to analyze the evolution of baryon density inhomogeneities.

\subsection{Time evolution}

\begin{figure}[ht] \centering 
\resizebox{80mm}{!}{\includegraphics{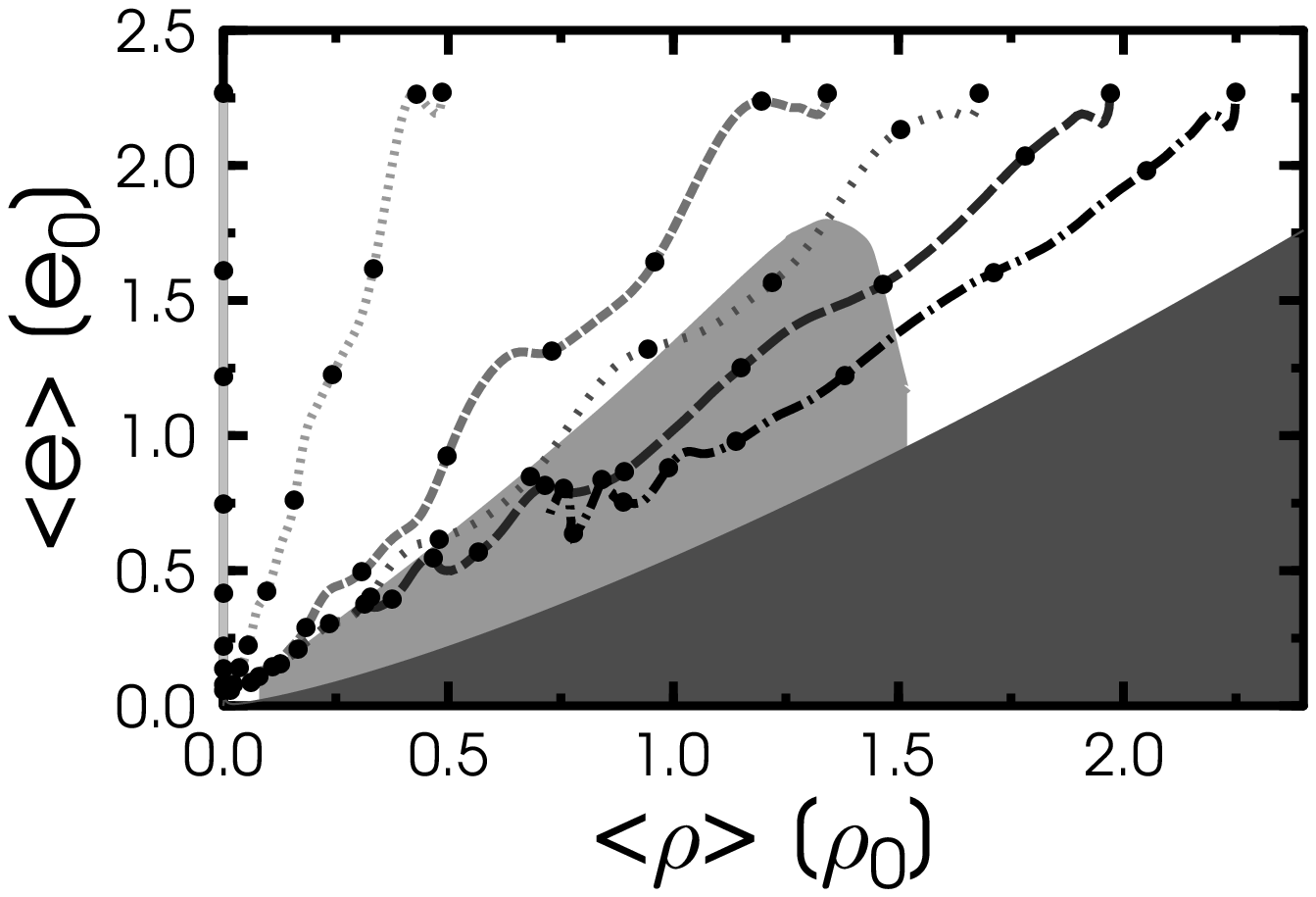}} 
\resizebox{80mm}{!}{\includegraphics{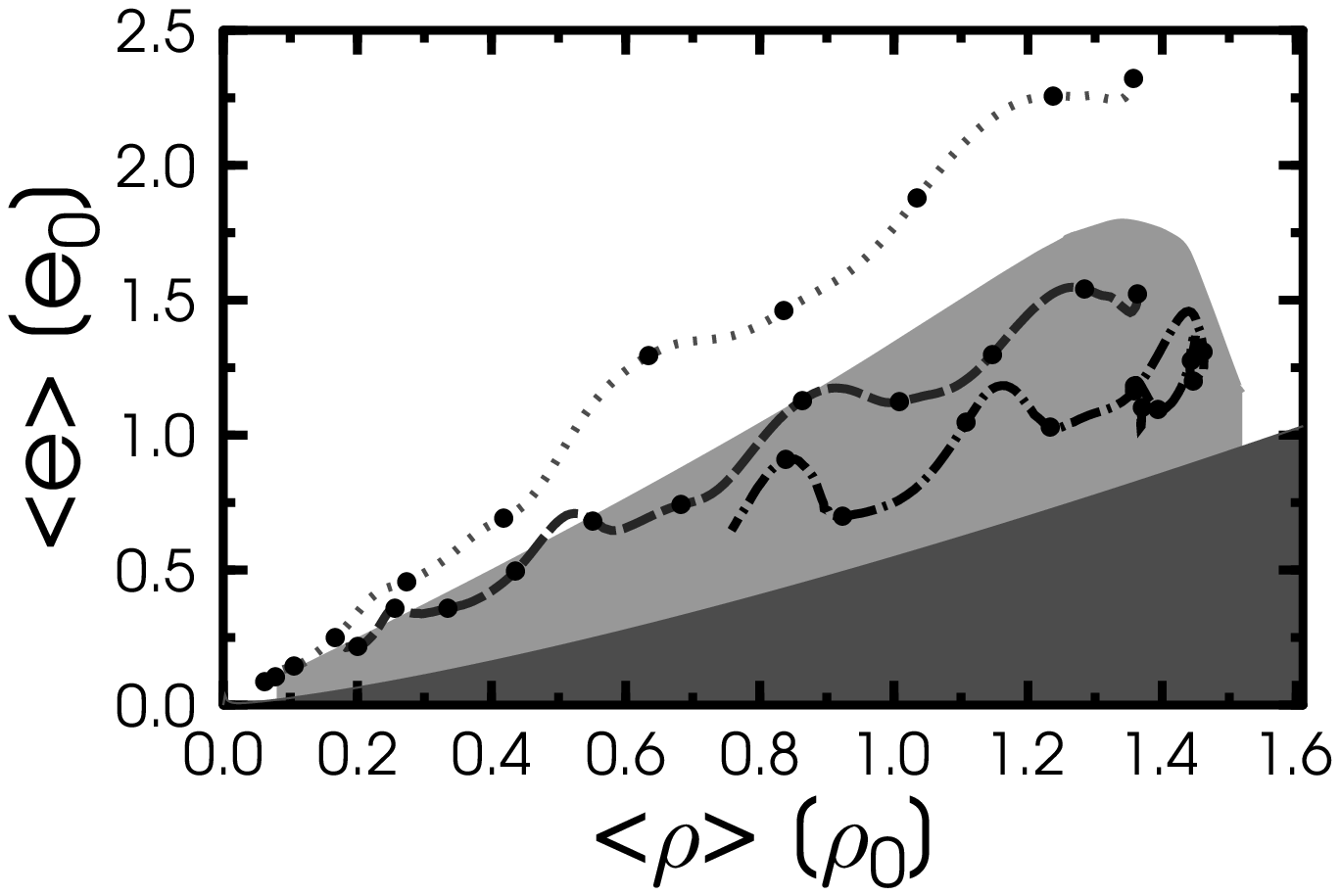}} 
\caption{Evolution of the average fluid energy and baryon density 
through a crossover, and a weak and strong first order
transition, respectively, for initial conditions I (top) and II (bottom).
The densities are measured in
  units of nuclear matter saturation density $\rho_0\approx
  0.16$~fm$^{-3}$, $e_0=m_N\rho_0\approx 0.15$~GeV/fm$^{3}$, with
  $m_N\approx0.922$~GeV the mass of a nucleon bound in infinite matter. 
The fat dots indicate time intervals of $\approx1.5$~fm/c.
The phase coexistence region is shaded in grey; the dark area depicts
the thermodynamically forbidden region where $e<\mu_B\rho-p$. } 
\label{e_rho_all} 
\end{figure} 
Fig.~\ref{e_rho_all} shows the trajectory of the system within the phase
diagram for both sets of initial conditions. 
For simplicity, we chose a foliation of space-time by flat
hypersurfaces without extrinsic curvature (i.e.\ surfaces of constant
CM-time). The average energy density of the quark fluid on such
surfaces is then given by
\be
\langle {e} \rangle(t) = \frac{ \int {\rm d}\sigma_\mu T^{\mu 0}_{\rm fluid}
  \,u_\sigma u_\nu T^{\sigma\nu}_{\rm fluid}}{
  \int {\rm d}\sigma_\mu T^{\mu 0}_{\rm fluid}}=
\frac{ \int {\rm d}^3x \;T^{0 0}_{\rm fluid} \,{e} }{
  \int {\rm d}^3x \;T^{0 0}_{\rm fluid}}~,
\ee 
and similarly for $\langle\rho\rangle$. We also average over several
initial field configurations picked according to~(\ref{FieldFlucs}).

The initial condition with $\mu_0<\mu_E$ evolves smoothly through a crossover.
For the other intitial conditions, 
the system enters the region corresponding to phase coexistence in the
equilibrium phase diagram and so undergoes a first order
phase transition. The explicit treatment of the dynamics of
the chiral fields (in the classical approximation) in space-time allows for
non-equilibrium effects and formation of inhomogeneities. 

Next, we determine the RMS fluctuation of the fluid density,
$\Delta\rho$, induced by the propagation of the Gaussian initial field
fluctuations~(\ref{FieldFlucs}) through the phase transition. First,
we determine the underlying smooth density profile on each
$t={\rm const.}$ time slice by averaging over the surface of a sphere
with thickness $\Delta r\simeq1$~fm,
\be \label{rho_profile}
\langle \rho\rangle(t,r) = {\cal N}^{-1}
 \int {\rm d}^3x \; \Theta(r+\Delta r-|\vec{x}|)\;
 \Theta(|\vec{x}|-r) \; \rho(t,\vec x)~,
\ee
with ${\cal N}=4\pi(\Delta r \;r^2 + \Delta r^{\;2}\;r + \Delta r^{\; 3}/3)$.
This profile is determined for each initial field configuration
individually. While averaging $\langle \rho\rangle(t,r)$ over
``events'' (i.e.\ initial field configurations),
too, would lead to seemingly larger $\Delta e$ and $\Delta\rho$, we
are interested here in density perturbations on scales of order 1~fm
within individual events. 

On each time slice, we
then define $\Delta \rho$ as the RMS deviation from this coarse-grained
density profile,
\begin{equation}
\Delta  {\rho}^{\;2} (t)= 
\frac{
\int{\rm d}^3x \left[ \rho (t,\vec{x}) -\langle \rho\rangle (t,r) \right]^2
\cdot \langle {e}\rangle (t,r)
}%
{%
\int{\rm d}^3x\; \langle {e}\rangle (t,r)~.
}
\end{equation}
We have chosen $\langle {e}\rangle (t,r)$, which is obtained in a
similar way as $\langle {\rho}\rangle (t,r)$ defined
in~(\ref{rho_profile}), as a weight in the integral 
to put more emphasis on the dense regions. Weighting with $\langle
{\rho}\rangle (t,r)$ instead leads to qualitatively similar results.

\begin{figure}[ht] 
\centering
\resizebox{80mm}{!}{\includegraphics{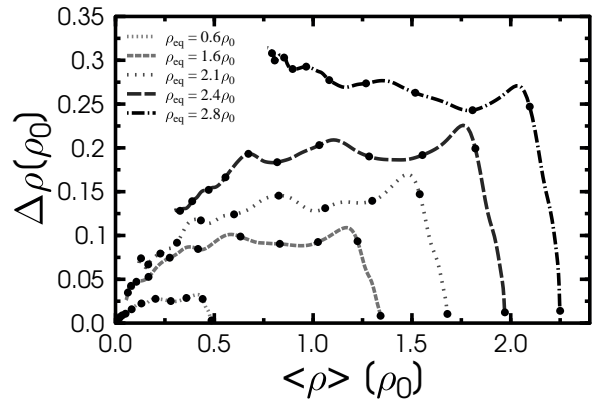}}
\caption{RMS fluctuations of the baryon
density with initial condition set (I) for crossover (narrow-dots, short-dashes),
weak (wide-dots, long-dashes) and strong 
(dash-dots) first order transitions as a function of the average
baryon density. The fat dots indicate time intervals of $\approx 1.5$~fm/c.
} 
\label{dx_all} 
\end{figure}
The time evolution of the baryon density inhomogeneities for initial
condition set (I)
is shown in fig.~\ref{dx_all}. We start with  fluctuations
of the order parameter field only, so that 
initially $\Delta e = \Delta  {\rho}=0$; this is to show the minimal
degree of inhomogeneity induced just by the transition to the symmetry
broken state.
As the evolution progresses, the fluctuations of the 
order parameter field rapidly lead to density inhomogeneities in the 
quark fluid.

One observes that the baryon density inhomogeneities are sensitive to
the dynamical evolution. The energy density inhomogeneities in this
model are smaller and show a weaker dependence on the initial baryon
density and are therefore not shown here.

For large initial baryon density the expansion
proceeds through the region of first-order phase transitions.
Here, the effective potential exhibits two
local minima within the ``phase coexistence'' region of the
equilibrium phase diagram (see e.g.\ fig.~1 in~\cite{Paech:2003fe} or
figs.~2-4 in~\cite{SMMR}) and so in some region of space the order
parameter can be ``trapped'' in the symmetric phase
until reaching the spinodal instability~\cite{ove}. This effect is
more pronounced the stronger the first-order phase transition, i.e.\
the smaller the entropy per baryon. Consequently,
density perturbations can only wash out after the double-minimum
structure of the effective potential has disappeared and the order
parameter ``rolls down'' to its new vacuum.
There is therefore reasonable hope that
these inhomogeneities created during the non-equilibrium phase
transition are present in the final state, contrary to those from
the initial state.
However, even for a crossover substantial
inhomogeneities could be present in the final state if they ``freeze''
shortly after passing the point where $V_{\rm eff}$ is flattest (or
where the chiral susceptibility peaks, respectively).

\begin{figure}[ht] 
\centering
\resizebox{80mm}{!}{\includegraphics{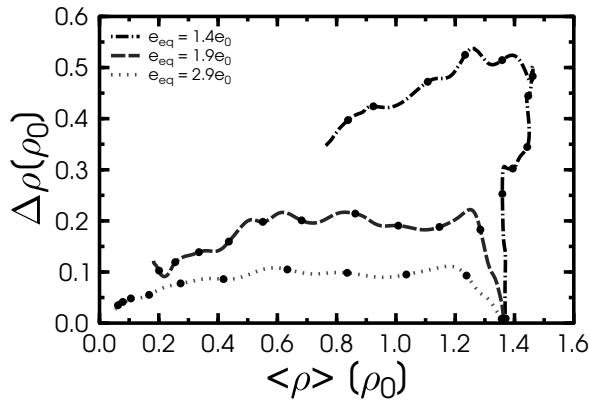}}
\caption{RMS fluctuations of the baryon
density with initial condition set (II) for crossover (dots), 
weak (dashes) and strong (dash-dots) first order transition
as a function of the average baryon density. 
The fat dots indicate time intervals of $\approx 1.5$~fm/c.} 
\label{dx_all_rhoconst} 
\end{figure}
Fig~\ref{dx_all_rhoconst} shows our results for the set (II) of
initial conditions, corresponding to 
{\em fixed} initial baryon density $\rho_{\rm eq}=1.7\rho_0$
but {\em different} initial energy density $e_{\rm eq}$. 
Again, the amplitude of the density contrast
is substantialy larger for a strong first order transition 
($e_{\rm eq}=1.4 e_0$) than for a crossover ($e_{\rm eq}=2.9 e_0$).

\section{Discussion} 

We have shown that the non-equilibrium dynamics of the order parameter
field in heavy ion collisions can lead to large density inhomogeneities
on the order of $\Delta \rho/\rho_0 \sim 0.1 -
1$. Further, that the 
amplitude of the fluctuations depends on the structure of the
effective potential: the effect
is stronger for a first-order phase transition than for a crossover.

What kind of experimental signatures could arise? 
By analogy to inhomogeneous Big Bang nucleosynthesis~\cite{ibbn},
which indeed is sensitive to
fluctuations of the baryon to photon ratio, one might expect that the
relative hadron abundances in heavy ion collisions are modified, too.
This is because the densities of various hadron species depend
non-linearly on the energy- and baryon density of the hadron
fluid and so fluctuations do not average out. 

The present model is too schematic to allow for quantitative
predictions of particle production. Nevertheless, experimental data
for relative hadron multiplicities could be analyzed, for example,
within the following simple model for an inhomogeneous decoupling
surface to test for the presence of (energy-) density
inhomogeneities. Thermal model fits to measured particle abundances
are commonly performed within the grand canonical ensemble, where the
density of any hadron species $i$ can be expressed in terms of the
temperature $T$ and the baryon-chemical potential $\mu_B$. Usually, a
uniform temperature and baryon-chemical potential is
assumed~\cite{chem_fits}. On the other hand, to test for
inhomogeneities, $T$ and $\mu_B$ could be taken as Gaussian random
variables. The average density of species
$i\in\left\{\pi,K,N,...\right\}$ is then given by
\bea 
\overline{\rho_i}\; (\overline{T},\overline{\mu}_B, \Delta
T,\Delta\mu_B) &=& \int\limits_0^\infty dT \; P(T;\overline{T},\Delta
T) \nonumber\\ 
& &\hspace{-3cm}\times \int\limits_{-\infty}^\infty
d\mu_B \; P(\mu_B; \overline{\mu}_B,\Delta\mu_B)~\rho_i (T,\mu_B)~,
\label{TmuDistrib}
\eea 
with $\rho_i(T,\mu_B)$ the actual ``local'' density of species
$i$ on the decoupling surface and 
\be 
P(x; \overline{x},\Delta x) \sim
\exp~ -\frac{\left(x-\overline{x}\right)^2}{2\; \Delta x^2} 
\ee 
the distribution of temperatures and chemical potentials.  The essential
point is that $\overline{\rho_i}\; (\overline{T},\overline{\mu}_B,
\Delta T,\Delta\mu_B) \neq \rho_i(\overline{T},\overline{\mu}_B)$ if
$\Delta T$, $\Delta\mu_B\neq0$.  
The main contribution to the integrals in~(\ref{TmuDistrib}) is not
from $\overline{T}$ and $\overline{\mu}_B$ but from the stationary
point of the integrand. For rare and heavy particles, where
quantum-statistical and relativistic effects can be neglected, there
is an exponential enhancement of the density with
$(\Delta\mu_B/\overline{T})^2$ and $(\Delta T/\overline{T})^2$; this
can be shown by a saddle-point integration
of~(\ref{TmuDistrib})~\cite{DPZinprep}.

{}From our results presented in the
previous section we expect that $\Delta T$, $\Delta\mu_B$ should be
significantly larger than zero if decoupling occurs close to the
first-order phase transition boundary. They should be smaller,
perhaps nearly zero, when the dynamical trajectory did not cross the
phase transition line. Hadron abundances at RHIC and SPS energies
could be studied within such a model to search for the presence of
inhomogeneities and to analyze their energy
dependence~\cite{detlef}. In fact, central $Au+Au$ collisions at top
AGS energy produce relatively cool but very baryon-dense
matter~\cite{HS} and could also probe the phase transition
line~\cite{AGS_pt}.  Other observables should also exhibit some
sensitivity to inhomogeneities, e.g.\ Hanbury-Brown--Twiss
correlations for pions~\cite{Socolowski:2004hw} or production cross
sections for light (anti-)~nuclei, formed by coalescence of
(anti-)~nucleons~\cite{yoffe}.

\acknowledgments We thank C.~Greiner for helpful discussions and for
reading the manuscript before publication and H.~J.~Drescher for
discussions on nucleon distributions in ground-state nuclei and
initial-state energy density fluctuations~\cite{comment}.\\
K.P.\ gratefully acknowledges support by
GSI. Numerical computations have been performed at
the Frankfurt Center for Scientific Computing (CSC).


\providecommand{\href}[2]{#2}\begingroup\raggedright

\endgroup

\end{document}